\documentclass[a4paper,twoside,notoc,11pt]{JHEP3}

\usepackage{epsfig,multicol,amsmath}

\newcommand\fverb{\setbox\pippobox=\hbox\bgroup\verb}
\newcommand\fverbdo{\egroup\medskip\noindent%
                            \fbox{\unhbox\pippobox}\ }
\newcommand\fverbit{\egroup\item[\fbox{\unhbox\pippobox}]}
\newbox\pippobox

\newcommand{\ie}{{\it i.e.}}

\newcommand{\cf}{{\it {cf. }}}
\newcommand{\wrt}{{\it {wrt. }}}

\newcommand{\bs}{\boldsymbol}
\newcommand{\qu}{{\rm q}}
\newcommand{\qb}{${\rm\bar q}$}
\newcommand{\qbm}{{\rm\bar q}}
\newcommand{\qq}{\qu\qb\ }

\newcommand{\rvec}{\vec r}
\newcommand{\Rvec}{\vec R}
\newcommand{\lqcd}{\Lambda_{QCD}}
\newcommand{\ieps}{i\varepsilon}

\newcommand{\eq}[1]{(\ref{#1})}

\newcommand{\beq}{\begin{equation}}
\newcommand{\eeq}{\end{equation}}
\newcommand{\beqa}{\begin{eqnarray}}
\newcommand{\eeqa}{\end{eqnarray}}
\newcommand{\nn}{\nonumber}

\renewcommand\jetp[3]{{\it Sov.\ Phys.\ JETP\/ }{\bf #1},{ #2}{ (#3)}}
\renewcommand\npb[3]{{\it Nucl.\ Phys.\ }{\bf B#1},{ #2}{ (#3)}}
\renewcommand\plb[3]{{\it Phys.\ Lett.\ }{\bf B#1},{ #2}{ (#3)}}

\renewcommand\prd[3]{{\it Phys.\ Rev.\ }{\bf D#1},{ #2}{ (#3)}}
\renewcommand\prep[3]{{\it Phys.\ Rept.\ }{\bf #1},{ #2}{ (#3)}}
\renewcommand\prl[3]{{\it Phys.\ Rev.\ Lett.\ }{\bf #1},{ #2}{ (#3)}}

\title{The Unexpected Role of Final State Interactions in Deep Inelastic Scattering\footnote{\uppercase{I}nvited talk at the {\it \uppercase{I}nternational \uppercase{C}onference: \uppercase{I}. \uppercase{Y}a. \uppercase{P}omeranchuk and \uppercase{P}hysics at the \uppercase{T}urn of the \uppercase{C}enturies,} \uppercase{M}oscow, \uppercase{J}anuary 2003. To be published in the proceedings.}}

\author{Paul Hoyer\thanks{Research supported in part by the
European Commission under contract HPRN-CT-2000-00130.}\\
            Department of Physical Sciences and Helsinki Institute of
            Physics\\
            POB 64, FIN-00014 University of Helsinki, Finland \\
            E-mail: \email{paul.hoyer@helsinki.fi}}
\preprint{HIP-2003-38/TH \\ \hepph{0307263}} 

\abstract{Rescattering of the struck quark in Deep Inelastic Scattering implies that measured parton distributions are not directly related to the Fock state probabilities of the target wave function. The production amplitudes acquire dynamical phases, which gives rise to shadowing and diffraction in DIS. I review the kinematics and dynamics of DIS as seen in various frames and gauges.}

\begin{document}

\section{Introduction}

Deep Inelastic Scattering, $e+ N \to e + X$ (Fig.~1) is one of our most precise tools for investigating the substructure of hadrons and nuclei. Soon after the first observation of scaling at SLAC, it was shown \cite{dy} that the DIS structure functions can be interpreted as the probability density of constituents in the target. Thus, in terms of the Fock state wave functions $\psi_n$ of a target nucleon $N$, DIS would measure the {\em parton density}
\beq 
\mathcal{P}_{\qu/N}(x_B,Q^2)= \sum_n
\int^{k_{i\perp}^2<Q^2}\left[ \prod_i\, dx_i\, d^2k_{\perp
i}\right] |\psi_n(x_i,k_{\perp i})|^2 \sum_{j=q} \delta(x_B-x_j)
\label{dens} 
\eeq 
The wave functions are to be evaluated at equal Light-Cone (LC) time $x^+=t+z/c$, and depend on the momentum fractions $x_i=k_i^+/p^+$ and transverse momenta $k_{\perp i}$ of the partons. The study of Ref.~1 predated QCD, and was based on a model of pseudoscalar meson - nucleon interactions. Later studies \cite{css} of QCD established the {\em factorization theorem} according to which the DIS cross section is given by the gauge invariant {\em parton distributions}
\beqa
f_{\qu/N}(x_B,Q^2)&=& \frac{1}{8\pi} \int dx^- \exp(-ix_B p^+ x^-/2)
\label{melm}\\
&\times&\langle N(p)| \qbm(x^-) \gamma^+\, {\rm P}\exp\left[ig\int_0^{x^-}
dw^- A^+(w^-) \right] \qu(0)|N(p)\rangle \nonumber
\eeqa
All fields are evaluated at a relative LC time $x^+ \sim 1/\nu$ and
transverse separation $x_\perp \sim 1/Q$, which vanish in the Bj limit (\cf Fig.~1).

\begin{figure}[bt]
\centerline{\epsfxsize=5cm\epsfbox{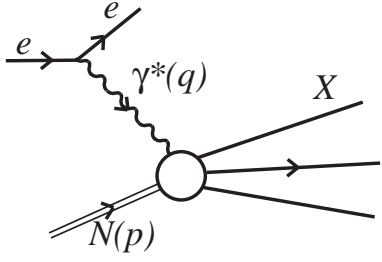}}   
\caption{Deep Inelastic Scattering (DIS), $eN \to eX$. The Bj limit is defined by $Q^2 \equiv -q^2 \to\infty$, $q\cdot p \equiv m_N\nu \to \infty$ with $x_B \equiv Q^2/2m_N\nu$ fixed.
 \label{fig1}}
\end{figure}

It would appear that the parton distribution \eq{melm} gives the parton density \eq{dens} in LC gauge ($A^+=0$), since the path ordered exponential is unity and the matrix element reduces to an overlap between Fock wave functions \cite{spd}. In this talk I shall discuss why the situation is not that simple in a gauge theory like QCD \cite{css,bhmps}. Complications arise because the LC gauge propagator 
\beq 
d_{LC}^{\mu\nu}(k) =
\frac{i}{k^2+\ieps}\left[-g^{\mu\nu}+\frac{n^\mu k^\nu+ k^\mu
n^\nu}{n\cdot k}\right]  \label{lcprop} 
\eeq
has a pole at $n\cdot k = k^+ =0$. Coulomb scattering of the struck quark on its way out of the target involves momentum transfers $k^+ \sim 1/\nu$ which makes LC gauge singular in the Bj limit.

Physically, the path ordered exponential in \eq{melm} describes soft rescattering of the struck (current) quark in the target. The quark has an asymptotically large momentum $p_\qu^- \sim \nu$ in the target rest frame and thus exits the target in an infinitesimal LC time $x^+ \sim 1/\nu$. The emission or absorption of physical, transversely polarized gluons requires a Lorentz-dilated formation time and thus cannot occur inside the target. Scattering from the target Coulomb field is, however, instantaneous and occurs over the whole traversal length (several fermis for a nuclear target). This is the reason that only the $A^+$ component of the target field appears in the leading twist matrix element \eq{melm}. It implies that the struck quark merely ``samples'' the target color field along its path, but does not influence it. Thus the matrix element is determined by the target wave function alone, as required by the QCD factorization theorem \cite{css}.

Coulomb rescattering affects the DIS cross section within the coherence (or ``Ioffe'') length $L_I$ of the hard scattering. In the target rest frame the coherence length of the virtual photon is $1/Q$ times the Lorentz factor $\nu/Q$,
\beq \label{ioffe}
L_I = \frac{1}{Q}\cdot\frac{\nu}{Q}= \frac{1}{2m_Nx_B}
\eeq
The Fourier transform in \eq{melm} shows that the parton distribution is indeed sensitive to Coulomb rescattering only over a distance $x^- \propto 1/m_Nx_B$.

Rescattering in the final state has important physical effects since it gives rise to kinematically {\em on-shell intermediate states} and hence to complex phases and interference between the rescattering amplitudes. Quantum mechanical interference is crucial for understanding the diffraction, shadowing and polarization effects which are prominently observed in DIS. The nucleon Fock amplitudes $\psi_n$ in \eq{dens} have no dynamical phases since their intermediate states are off-shell.

It is thus intuitively plausible that rescattering is an essential part of DIS physics which cannot be turned off by a gauge choice. In the following I shall give a more detailed description of DIS dynamics based on our experience from a perturbative model \cite{bhmps}.

\section{Two views of DIS from the target rest frame}

According to \eq{ioffe} the coherence length is long at at low $x_B$. This is also the region where shadowing and diffraction phenomena set in. In the following I shall therefore have low $x_B$ in mind -- but the general conclusions are valid at any $x_B$. 

There are two essentially different frames for viewing DIS dynamics, often referred to as the ``Infinite Momentum Frame'' and ``Target Rest Frame'' (see \cite{Collins:2001hp} for a discussion). Actually both views may be had in the target rest frame (LC physics is invariant under longitudinal boosts). The crucial difference is whether the photon moves in the negative or positive $z$-direction. The description in terms of target LC wave functions defined at equal LC time $x^+$ is quite different in the two cases.

\subsection{DIS frame}

I shall refer to the target rest frame where the photon moves along the negative $z$-axis as the {\em DIS frame}: $q = (\nu,\bs{0}_\perp,-\sqrt{\nu^2+Q^2})$. The view in the DIS frame is equivalent to that in the Infinite Momentum and Breit frames. We have then 
\beq \label{qdis}
q^- \simeq 2\nu,\hspace{1cm} q^+ = -Q^2/q^- \simeq -m_N x_B <0\ \  \mbox{ and }\ \  q_\perp=0
\eeq
 Since the photon moves essentially with the velocity of light, it traverses the target in vanishing LC time, $x^+ = t+z \simeq 1/\nu$, even though both $t$ and $z$ are of $\mathcal{O}$(fm). This is why the DIS frame gives a snapshot of the proton wave function at a given $x^+$.

In LC time ordered perturbation theory (LCPTH) all 4-momenta are treated as on-shell, and the ``plus'' components are positive\footnote{The virtual photon is here regarded as an external particle with negative squared mass $q^2=-Q^2$ and thus, exceptionally, $q^+ < 0$ in \eq{qdis}.}. The conservation of ``plus'' momentum then forbids the transition $\gamma^*(q) \to \qu\qbm$ for $q^+<0$. Thus we have only the LC time ordering $\gamma^*\qu \to \qu$: the photon scatters on a target quark, revealing target structure. Space-time coherence in the DIS frame is governed by
\beqa \label{discoh}
x^+ &\sim& \frac{1}{q^-}\simeq \frac{1}{2\nu}\to 0 \makebox[7.7cm][r]{Target wf is probed at equal LC time}
\eeqa

An $x^+$-ordered picture of the scattering is shown in Fig.~2a. The rescattering of the struck quark on the gluon field of the target influences the DIS cross section insofar as it occurs within the Ioffe coherence length \eq{ioffe}. Even though the rescattering occurs over a finite distance in time and space the LC time is infinitesimal, $x^+ \sim 1/\nu$ as given in \eq{discoh}. Hence one may regard the rescattering effects as part of an ``augmented'' LC wave function. Such a wave function contains physics specific to the DIS probe, and is thus distinct from the usual wave function of an isolated hadron.

\begin{figure}[bt]
\centerline{\epsfxsize=10cm\epsfbox{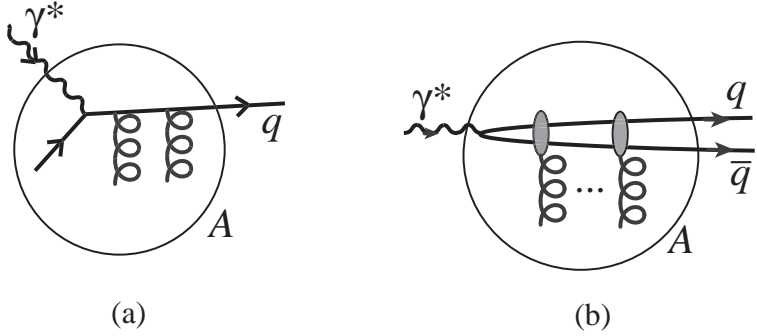}}   
\caption{LC time $(x^+=t+z)$ ordered dynamics of deep inelastic scattering. (a) The DIS frame (2.1), where the virtual photon momentum $q^z \simeq -\nu$. The photon hits a target quark, which Coulomb rescatters before exiting the target $A$. The increase in $t$ is compensated by a decrease of $z$ such that the photon probes the target at an instant of $x^+$. (b) The Dipole frame (2.3), where $q^z \simeq +\nu$. The photon splits into a \qq pair at LC time $x^+ \sim 1/2m_N x_B$ before the target. The DIS cross section is given by the scattering of the \qq color dipole in the target.
 \label{fig2}}
\end{figure}

\subsection{Dipole frame}

I shall refer to the target rest frame where the photon moves along the positive $z$-axis as the {\em Dipole frame}: $q = (\nu,\bs{0}_\perp,+\sqrt{\nu^2+Q^2})$. This frame is related to the DIS frame by a $180^{\circ}$ rotation around a transverse axis. Such a rotation is a dynamical transformation for a theory quantized at $x^+=0$. Hence there is no simple relation between the DIS and dipole frames in the case of non-perturbative target structure (the two frames can obviously be connected for a perturbative target model). The kinematics is inverted \wrt that of \eq{qdis} and \eq{discoh},
\beqa \label{dipcoh}
q^+ \simeq 2\nu \hspace{2cm}&& x^- \sim \frac{1}{2\nu} \to 0 \nn \\
q^- \simeq -\frac{Q^2}{2\nu} \hspace{2cm} && x^+ \sim \frac{1}{m_N x_B} 
\eeqa
Hence in the dipole frame the scattering occurs over a finite LC time $x^+ \sim L_I$. Since $q^+$ is positive there can be a $\gamma^* \to \qu\qbm$ transition before the interaction in the target. The $\qu\qbm$ forms a color dipole whose target cross section is determined by its transverse size. In the aligned jet model -- which corresponds to the parton model and thus to lowest order in QCD -- the quark takes nearly all the longitudinal momentum, $p_\qu^+ \simeq q^+ \simeq 2\nu$, while $p_\qbm^+ \sim \lqcd^2/m_Nx_B$. The transverse separation of the quark pair is then $r_\perp(\qu\qbm) \sim 1/\lqcd \sim 1$ fm. Such a transversally large \qu\qb\ dipole has a non-perturbative cross section, which is the parameter in the dipole frame that corresponds to the parton distribution of the DIS frame.

The \qu\qb\ pair multiple scatters in the target as shown in Fig.~2b. At low $x_B$ also the antiquark momentum is large and one expects pomeron exchange, \ie, diffractive DIS, as well as shadowing of the target \cite{gribov}. The dipole frame is natural for modelling these features of the data, where phases and interference effects play a key role \cite{Piller:1999wx}. 

If it were possible to ``turn off'' rescattering in the DIS frame by choosing LC gauge $A^+=0$, the pomeron and interference effects of the dipole frame would have to be built into the target wave function itself (which, as I noted above, has no dynamical phases). Thus there would be a preformed pomeron in the target, and shadowing would imply that quarks with low $x_B$ would be suppressed in nuclei (relative to nucleons).

I shall next discuss why rescattering effects cannot be avoided even in the DIS frame. This makes the physics of the DIS and dipole frames much more similar. In particular, diffraction and shadowing arise from the interference of rescattering amplitudes in both frames. 

\begin{figure}[bt]
\centerline{\epsfxsize=8cm\epsfbox{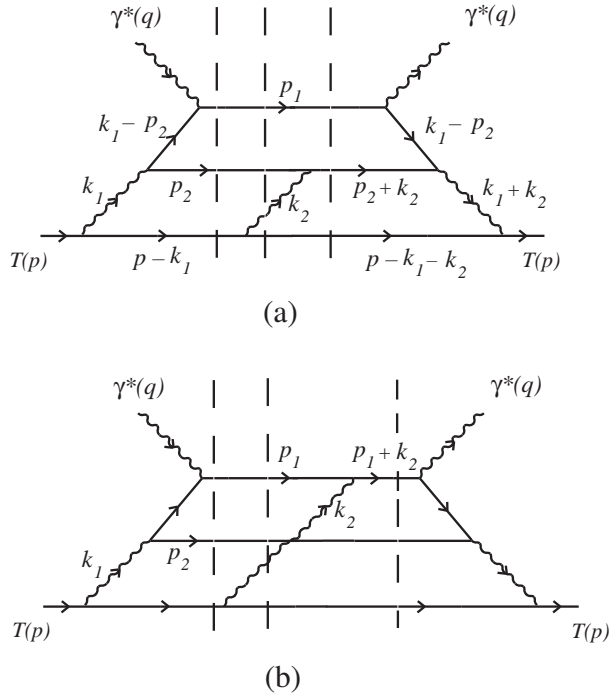}}   
\caption{Two types of final state interactions. (a) Scattering of the
antiquark ($p_2$ line), which in the aligned jet kinematics is part of the
target dynamics. (b) Scattering of the current quark ($p_1$ line). For each LC
time-ordered diagram, the potentially on-shell intermediate states are marked by dashed
lines. \label{fig3}}
\end{figure}

\section{Interactions between spectators}

We normally regard the target as ``frozen'' during the DIS process. I first recall why this is legitimate in Feynman gauge, and then discuss why there are relevant spectator interactions in LC gauge.

\subsection{Spectator interactions in Feynman gauge}

The generic, $x^+$-ordered Feynman diagram shown in Fig.~3a involves a gluon exchange between two target spectator lines (in the aligned jet kinematics only the struck quark carries asymptotically large momentum, $p_1^- \simeq 2\nu$ in the target rest frame). Three intermediate states, marked by dashed lines, can kinematically be on-shell and thus contribute to the discontinuity of this forward diagram. Since $p_1^-$ is much bigger than all target momenta the three discontinuities occur at essentially the same value of $p_1^-$. Hence they form a higher order pole, \ie, their contributions cancel to leading order in the Bj limit. The same argument applies to any diagram involving spectator-spectator exchanges between the virtual photon vertices.

Exchanges between a spectator and the struck quark (Fig.~3b) are, on the other hand, expected to contribute at leading twist and to build the path-ordered exponential in \eq{melm}. Elastic scattering of the struck quark implies $k^+ \sim 1/\nu$, which is of the same order as $p^+_1$. There are then two distinct large momenta in Fig.~3b, $p_1^-$ and $(p_1+k_2)^-$, and the three discontinuities do not cancel.

The above argument holds in Feynman gauge, where all contributions to the discontinuity arise from on-shell configurations.

\subsection{Spectator interactions in LC gauge}

The singularity of the LC gauge propagator \eq{lcprop} at $k^+=0$ blurs the distinction between spectator-spectator and spectator-struck quark interactions (Figs.~3a and 3b, respectively). Since the $k^+=0$ poles are gauge artifacts the sum of their residues must add up to zero. However, this cancellation requires that diagrams of type 3a and 3b are added.

The path ordered exponential in \eq{melm} reduces to unity in $A^+=0$ gauge. In perturbation theory this occurs through a cancellation\footnote{Depending on the prescription used at $k^+=0$ this cancellation can occur separately for each Feynman diagram, or only in their combined contribution to the DIS cross section, as explained in Ref. \cite{Belitsky:2002sm}.} between the $-g^{\mu\nu}$ (Feynman) and $(n^\mu k^\nu+k^\mu n^\nu)/k^+$ (LC gauge artifact) parts of the propagator \eq{lcprop}, for interactions of the struck quark such as in Fig.~3b.

Once the $k^+=0$ poles of the LC gauge propagator are ``used up'' in diagrams like Fig.~3b to cancel the Feynman gauge contribution, the corresponding spurious poles in diagrams like Fig.~3a give a non-vanishing contribution to the leading twist DIS cross section. In fact, their sum must equal the contribution from the path ordered exponential in Feynman gauge. Thus gauge independence is achieved, and the simplification of the exponential ({\it alias} rescattering of the struck quark) is accompanied by a complication in the spectator system.

\begin{figure}[bt]
\centerline{\epsfxsize=14cm\epsfbox{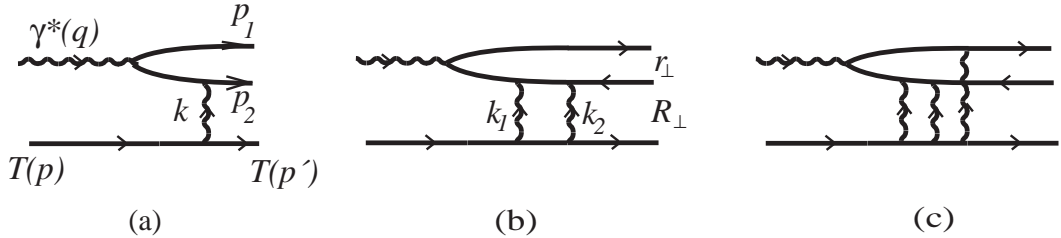}}   
\caption{A scalar abelian model for deep inelastic scattering with one-, two, and three-gluon exchanges. At each order only one representative diagram is shown. In the $x_B \to 0$ limit and at the orders considered no other final states contribute to the total DIS cross section. \label{fig4}}
\end{figure}

\section{A perturbative model}

Most of the features discussed above were verified in an perturbative model calculation \cite{bhmps} (\cf Fig. 4). The one-, two- and three-gluon exchange amplitudes were evaluated in the Bj limit for $x_B \to 0$ in both Feynman and LC gauge, using various prescriptions for the $k^+ = 0$ poles, and in covariant as well as $x^+$-ordered perturbation theory. In this way the gauge dependence of the individual diagrams could be seen to be compatible with the gauge independence of the physical cross section. In particular, the fact that the $k^+ = 0$ poles are absent in the complete sum of diagrams was verified.

In impact parameter space the $n$-gluon exchange amplitudes were (for $n=1,2,3$) found to have the form $A_n \propto V(\gamma^*)W^n$, where $V$ is the virtual photon wave function and $W$ a rescattering factor of the \qq\ pair. This corresponds to the structure expected in the dipole frame at small $x_B$, where the quark pair is created by the photon and then rescatters in the target. The perturbative calculation can of course be equivalently done in the DIS frame. 

Summing the gluon exchanges to all orders gives
\beq
|\sum_{n=1}^\infty A_n| = 
\left|\frac{\sin \left[g^2\, W(\rvec_\perp,
\Rvec_\perp)/2\right]}{g^2\, W(\rvec_\perp,
\Rvec_\perp)/2} A_1(p_2^-,\rvec_\perp, \Rvec_\perp)\right|
\label{Interference}
\eeq
where
\beq
W(\rvec_\perp, \Rvec_\perp) = \frac{1}{2\pi}
\log\left(\frac{|\Rvec_\perp+\rvec_\perp|}{R_\perp} \right)
\label{Wexpr}
\eeq
with $\rvec_\perp$ the transverse size of the produced quark pair, $\Rvec_\perp$ the distance between the pair and the target (see Fig.~4b), and $g$ the gauge coupling. Hence the higher order corrections reduce the magnitude of the Born amplitude $A_1$, and thus also of the DIS cross section. This ``shadowing'' effect arises from destructive interference between the rescattering amplitudes $A_n$.

\begin{figure}[hbt]
\centerline{\epsfxsize=7cm\epsfbox{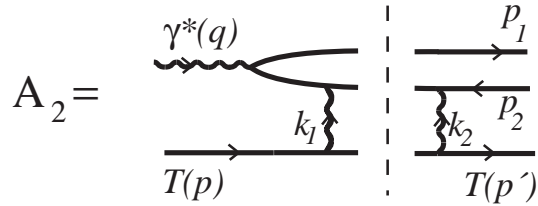}}   
\caption{The two-gluon exchange amplitude $A_2$ is purely imaginary at low $x_B$. The intermediate state indicated by the dashed line is thus on-shell and the full amplitude is given by the product of the two subamplitudes on either side of the cut. (Only one of the contributing Feynman diagrams is shown.)
\label{fig5}}
\end{figure}

The rescattering gives the amplitudes complex phases. While the Born ($A_1$) amplitude is real, the two-gluon exchange amplitude $A_2$ is purely imaginary in the $x_B \to 0$ limit, as mandated by analyticity and crossing. Hence the intermediate state between the gluon exchanges in Fig.~4b is on-shell. This allows to write the full amplitude as a product of two single gluon exchange subamplitudes (Fig.~5). The subamplitudes are on-shell and hence gauge invariant: The feature of on-shell intermediate states, and hence rescattering, is present in Feynman as well as in LC gauge.

The $A_2$ amplitude is also the lowest order contribution to diffractive DIS, characterized by color singlet exchange. This leading twist part of the total DIS cross section thus explicitly arises from rescattering of the struck quark in the target.

\begin{figure}[hbt]
\centerline{\epsfxsize=10cm\epsfbox{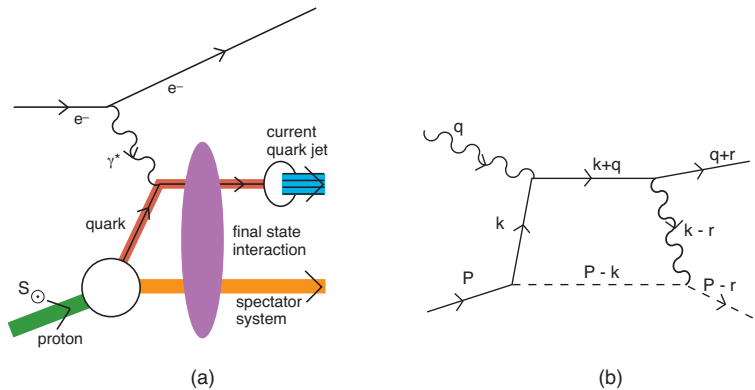}}   
\caption{(a) The struck quark can be asymmetrically distributed in azimuth wrt the virtual photon direction if it suffers final state interactions and the target is transversely polarized. (b) A model diagram with a final state interaction giving rise to a transverse spin asymmetry at leading twist. (The figures are from Ref. [9].)
\label{fig6}}
\end{figure}

\section{Single spin asymmetry}

The realization that rescattering gives rise to complex phases in DIS amplitudes led to interesting developments for the transverse spin asymmetry in DIS \cite{Brodsky:2002cx,Collins:2002kn}. The data \cite{Airapetian:1999tv} indicated a sizeable target spin asymmetry, in apparent conflict with a theorem \cite{Collins:1992kk} that such effects are higher twist. The asymmetry for a transversely polarized target is given by the imaginary part of the interference between target helicity flip and non-flip, $SSA \sim Im(T_\lambda T^*_{-\lambda})$, and is thus sensitive to phase differences. The gluon rescattering contribution in the model Feynman diagram of Fig.~6b provides such a phase, resulting in an $SSA$ at leading twist \cite{Brodsky:2002cx}.

The rescattering phases are contained in the path ordered exponential of the parton distribution \eq{melm}. The presence of the exponential violates the assumptions that were made \cite{Collins:1992kk} in deriving the theorem, and  indeed allows a leading twist asymmetry \cite{Collins:2002kn}. This effect is unrelated to the transverse spin carried by the struck quark, making a measurement of the latter more challenging.

\section{Discussion}

Our discussion has centered on why the effects of the path ordered integral
\beq
{\rm P}\exp\left[ig\int_0^{x^-} dw^- A^+(w^-) \right]
\eeq
in the expression \eq{melm} of the parton distribution cannot be eliminated. The exponential reduces to unity in light cone gauge ($A^+=0$), but this merely shifts its effects to the spectator system. The exponential arises from the rescattering of the struck quark in the target. This makes the scattering amplitudes complex, giving rise to diffraction and shadowing phenomena in DIS. Such physical effects cannot be eliminated by a gauge choice.

Some misunderstandings occur due to imprecise use of the term ``final state interactions''. In the present context we should differentiate between interactions at three time scales (in the target rest frame):
\begin{enumerate}
\item The hadronization time of the struck quark. Due to Lorentz dilation this is proportional to the quark energy: $t=x^0\propto \nu$. The total DIS cross section is independent of interactions occurring at the hadronization time scale.
\item The Ioffe time $t_I=\nu/Q^2 = 1/(2m_N x_B)$, which is finite in the Bj limit. Interactions occurring within the Ioffe time affect the DIS cross section. The struck quark rescattering which gives rise to the path ordered exponential occurs at this time scale. Note that the corresponding LC time $x^+ = t+z \propto 1/\nu$ is infinitesimal in the DIS frame, due to a cancellation between the time $t$ and the distance $z$ when the photon moves in the negative $z$-direction.
\item The exact LC time $x^+$ of the virtual photon interaction. This is relevant for QCD quantized at a given $x^+$, and in $x^+$-ordered perturbation theory. Interactions occurring before the photon interaction build the light cone wave function of the target. Struck quark rescattering occuring after this time gives rise to dynamic phases which are specific to DIS, and are not present in the wave function of an isolated hadron.
\end{enumerate}

The struck quark rescattering that occurs at the Ioffe time is within the time (or $x^+$) resolution of the virtual photon. It is therefore possible to regard those interactions as part of an ``augmented'' target wave function, specific to the hard inclusive probe. Since the interactions occur at infinitesimal $x^+\propto 1/\nu$ and the exchanged gluons likewise carry $k^+ \propto 1/\nu$, the augmented wave function apparently differs from the standard one through ``zero modes''. Such modes \cite{Brodsky:1997de} are possible for theories quantized on a light-like surface, which allows causal (light-like) connections.

It is sobering to realize that many questions are still open concerning such a basic QCD process as deep inelastic scattering, more than 30 years after it first was discovered.

\section*{Acknowledgments}

I would like to thank the organizers of this meeting for their kind invitation. The work I have presented is based on work done with Stan Brodsky, Nils Marchal, St\'ephane Peign\'e and Francesco Sannino. I am grateful to them and many others for helpful discussions.

\end{document}